\documentclass[12pt,preprint]{aastex}

\slugcomment{Accepted for publication in {\it The Astronomical Journal}}

\shorttitle{GRAPES/UDF Redshifts}
\shortauthors{Xu et al.}

\begin{document}

\def\oii{[\ion{O}{2}] $\lambda$3727}
\def\nii{[\ion{N}{2}] $\lambda$6583}
\def\oiii{[\ion{O}{3}] $\lambda$5007}
\def\oiiic{[\ion{O}{3}] $\lambda$4995}
\def\oiiipair{[\ion{O}{3}] $\lambda \lambda$4959,5007}
\def\civ{[\ion{C}{4}] $\lambda$1549}
\def\ciii{[\ion{C}{3}] $\lambda$1909}
\newcommand{\lya}{Lyman-$\alpha$}
\def\erg{\hbox{erg}}
\def\cm{\hbox{cm}}
\def\sec{\hbox{s}}
\def\ergcm2s{\ifmmode {\rm\,erg\,cm^{-2}\,s^{-1}}\else
                ${\rm\,ergs\,cm^{-2}\,s^{-1}}$\fi}
\def\ergsec{\ifmmode {\rm\,erg\,s^{-1}}\else
                ${\rm\,ergs\,s^{-1}}$\fi}
\def\kmsMpc{\ifmmode {\rm\,km\,s^{-1}\,Mpc^{-1}}\else
                ${\rm\,km\,s^{-1}\,Mpc^{-1}}$\fi}
\def\kms{\ifmmode {\rm\,km\,s^{-1}}\else
                ${\rm\,km\,s^{-1}}$\fi}
\def\lstar{\ifmmode {L_\star}\else
                ${L_\star}$\fi}
\def\phistar{\ifmmode {\phi_\star}\else
                ${\phi_\star}$\fi}
\def\Mpc{\hbox{Mpc}}
\def\Msun{M_{\odot}}
\def\mag{\ifmmode {\,\hbox{mag}}\else
                 mag\fi}

\title{Redshifts of Emission Line Objects in the Hubble Ultra Deep Field}

\author{Chun Xu\altaffilmark{1,2},
        Norbert Pirzkal\altaffilmark{1},
        Sangeeta Malhotra\altaffilmark{1,8},
        James E. Rhoads\altaffilmark{1,8}, 
        Bahram Mobasher\altaffilmark{1},
        Emanuele Daddi\altaffilmark{3},
        Caryl Gronwall\altaffilmark{4},
        Nimish P. Hathi\altaffilmark{8},
        Nino Panagia\altaffilmark{1},
        Henry C. Ferguson\altaffilmark{1},
        Anton M. Koekemoer\altaffilmark{1},
        Martin K\"ummel\altaffilmark{5},
        Leonidas A. Moustakas\altaffilmark{1},
        Anna Pasquali\altaffilmark{6},
        Sperello di Serego Alighieri\altaffilmark{7},
        Joel Vernet\altaffilmark{7},
        Jeremy R. Walsh\altaffilmark{5},
        Rogier Windhorst\altaffilmark{8},
        Haojing Yan\altaffilmark{9}
}

\altaffiltext{1}{Space Telescope Science Institute, 3700 San Martin Drive,
    Baltimore, MD 21218.}
\altaffiltext{2}{Email: cxu\_25 ``at'' yahoo.com}
\altaffiltext{3}{{\em Spitzer Fellow}, National Optical Astronomy 
 Observatory, P.O. Box 26732, Tucson, AZ 85726.}
\altaffiltext{4}{Department of Astronomy, Pennsylvania State
    University, University Park, PA 16802.}
\altaffiltext{5}{ESO/ST-ECF, Karl-Schwarzschild-Strasse 2, D-85748,
    Garching, Germany.}
\altaffiltext{6}{Institute of Astronomy, ETH Hoenggerberg, CH-8093
    Zurich, Switzerland.}
\altaffiltext{7}{INAF - Osservatorio Astrofisico di Arcetri, Largo
    E. Fermi 5, I-50125 Firenze, Italy.}
\altaffiltext{8}{School of Earth and Space Exploration,
    Arizona State University, Tempe, AZ 85287.}
\altaffiltext{9}{Spitzer Science Center, Caltech, MS 100-22, Pasadena, 
    CA 91125.}

\begin{abstract}
We present redshifts for 115 emission line objects in the
Hubble Ultra Deep Field (UDF) identified through the 
GRism ACS Program for Extragalactic Science (GRAPES) project
using the slitless grism spectroscopy mode 
of the ACS Camera on the Hubble Space Telescope (HST).  
The sample was selected by an emission line search on all
extracted 1-dimensional GRAPES spectra.
We identify the emission lines using line wavelength
ratios where multiple lines are detected in the
grism wavelength range ($5800\hbox{\AA} \la \lambda \la 9600$\AA),
and using photometric redshift information where multiple
lines are unavailable.  We then derive redshifts using the
identified lines.  Our redshifts are accurate to $\delta z \approx 0.009$,
based on both statistical uncertainty estimates and
comparison with published ground-based spectra.
Over 40\% of our sample is fainter than typical magnitude limits 
for ground-based spectroscopy (with $i_{AB}>25 \mag$). 
Such emission lines would likely remain undiscovered without
our deep survey. 
The emission line objects fall into 3 categories: 1) Most are
low to moderate redshift galaxies ($0 \le z \le 2$), including many
actively star forming galaxies with strong HII regions; 
2) 9 are high redshift ($4 \le z \le 7$) Lyman-$\alpha$ emitters;
and 3) at least 3 are candidate AGNs. 
\end{abstract}


\section{Introduction}

The GRism ACS Program for Extra-galactic Science (GRAPES)
project is a slitless spectroscopic survey of
the Hubble Ultra Deep Field (UDF) region (Beckwith 2006).
GRAPES exploits the high spatial
resolution and low background of the G800L grism on the
Advanced Camera for Surveys (ACS).  The resulting spectra
can detect continuum levels down to $z_{AB} > 27$ magnitude,
making them the deepest ever obtained for continuum spectroscopy.
Additionally, they can detect emission lines to $\sim 5\times 
10^{-18} \ergcm2s$, more sensitive than any previous slitless survey
and comparable to the sensitivities of typical slit
spectra on $6$--$10$ meter class telescopes.
GRAPES thereby provides a rich
data set with a wide range of potential applications,  from
Lyman breaks in distant galaxies (Malhotra et al. 2005), 
to 4000 \AA\ breaks in galaxies with older stellar populations (Daddi
et al. 2005; Pasquali et al. 2005), to spectral classification of
faint Galactic stars (Pirzkal et al. 2005).  

The use of slitless spectrographic surveys to search for emission-line
galaxies dates back more than three decades.  The most famous surveys
of this type used objective prisms on Schmidt telescopes.  Early surveys
utilized photographic plates in order to maximize the field-of-view.
Surveys of this type include the Tololo (Smith 1975; Smith et al. 1976)
and Michigan (MacAlpine et al. 1977, MacAlpine \& Williams 1981) surveys.
More recently the Universidad Complutense de Madrid (Zamorano et al.
1994, 1996) survey defined a well-studied sample of H$\alpha$ selected
galaxies.  The advent of large CCD detectors available on Schmidt
telescopes has made possible large-scale digital objective-prism surveys.
The first of this type is the KPNO International Spectroscopic Survey
(KISS; Salzer et al. 2000).  Objective prism surveys from the ground
have in general been most effective for relatively nearby galaxies.
Efforts to search for high redshift Lyman-$\alpha$ emitting galaxies
with slitless spectrographic surveys also be dated back to as early
as 1980s (e.g., Koo \& Kron 1980; for a review, see Pritchet 1994).
However, only the advent of slitless spectroscopic capabilities 
on the Hubble Space Telescope on both the STIS and NICMOS instruments, 
enabled such surveys to achieve high sensitivity. Examples include 
H$\alpha$ from redshift 
0.75 to 1.9 with NICMOS (McCarthy et al. 1999; Yan et al. 1999)
and \oii\ from 0.43 to 1.7 with STIS (Teplitz et al.
2003a,b). These surveys were done in parallel observing mode to maximize
the total area observed.  These were followed by parallel mode
surveys using ACS during the first year of ACS operations: The APPLES
survey, led by Rhoads (see Pasquali et al 2005),
and a similar effort led by L. Yan (see Drozdovsky et al 2005).
GRAPES is a natural successor to these efforts, and represents 
a major step forward in sensitivity and robustness, thanks to the
improved experimental design allowed by pointed observations.

In the present paper, we present a redshift catalog for strong emission
lines galaxies identified in the GRAPES spectra for both nearby and 
distant galaxies.  We combine our line wavelengths
with photometric redshift estimates from broad band photometry to
obtain accurate redshifts for most of the emission line sources in the
UDF.  Such redshifts are a starting point in studies of cosmological
evolution.  Emission line galaxies are of particular physical interest
for several reasons.  $H\alpha$, \oii\ and \oiiipair\ can be 
used to study the evolution of star formation rate (e.g., Gallego et
al. 1995, 2002).  We will present such analysis from the
GRAPES emission lines, including the completeness analysis,
in forthcoming paper (Gronwall et al., in preparation).
Ly$\alpha$ can be a prominent signpost of actively star forming
galaxies at the highest
presently available redshifts, and can be used to probe physical
conditions in these galaxies (Malhotra \& Rhoads 2002) and to study
the ionization state of the intergalactic medium (Malhotra \& Rhoads
2004, 2006).

The paper is organized as follows. We describe the data reduction
and emission line search procedure in section 2, then we present
our results in section 3, followed by a discussion
section (section 4).  We present our redshift catalog for 
emission line galaxies in table~1. 

\section{Observations and Reductions}
Slitless spectroscopy using the ACS G800L grism
on the Hubble Space Telescope has its own unique characteristics,
which we discuss briefly here as useful background to understanding
the strengths and limitations of our data set.
First, the high spatial resolution and low sky background
afforded by HST result in highly sensitive spectra.  Second,
the dispersion is low, at $40$\AA\ per pixel.  
Third, the slitless nature of the observations imply that spectra may 
overlap, and that the effective spectral resolution scales inversely 
with the angular size of the source.  Thus, to detect an emission
line, its equivalent width (in \AA) should not be far below
the object size times spectrograph dispersion.  Our
selection effects for a range of continuum magnitude, line flux,
and equivalent width are discussed below.  Fourth, continuum subtraction
is nontrivial in grism spectra.  The basic
difficulty is in identifying object-free regions of the dispersed
grism data to scale and subtract a sky model.  This can be done very well,
but not perfectly, with residuals at the level of a few $\times 10^{-4}$
counts/s/pixel, corresponding to $\sim 10^{-3}$ of the sky count rate
(see Pirzkal et al 2004 for further discussion).
The implication for emission line
studies is that equivalent width measurements at the faintest continuum
fluxes are subject to potential error induced by continuum subtraction
residuals.  Fifth, the grism signal-to-noise ratio and the contamination 
can both be helped by using a narrow extraction window, but this 
comes at the cost of aperture losses in the extracted spectrum.  
Because the ACS point spread function does not vary strongly
with wavelength, the aperture losses should be largely wavelength
independent, and will have little effect on line ratio measurements.

The GRAPES observations consisted of a total of 5 epochs of observations 
with comparable exposure time at 5 different orientations
(also quoted as Position Angles or PAs).  By splitting the
observations into different orientations, we are able
to mitigate the effects of overlapping spectra: most objects
are uncontaminated in at least some roll angles.
Due to slight offsets in sky coverage for different roll angles,
we have varied depth in the exposures. About 10.5 square arcmin
are covered by at least 4 roll angles, and more than 12.5 square arcmin
are covered by at least one roll angle.
Because the spectra are $\sim 100$ pixels long, while the
field size is $4096$ pixels, the fraction of spectra truncated
by the field edge (and thereby lost) is only $\sim 2\%$.
A more detailed
description of GRAPES project, especially the observations and
the data reduction, can be found in Pirzkal et al. (2004).
We note that the spectral resolution, line sensitivity,
and equivalent width threshold for GRAPES emission lines are all
comparable to typical modern narrowband surveys (e.g.,
Rhoads et al 2000, 2004; Rhoads \& Malhotra 2001; Malhotra \& Rhoads
2002; Cowie \& Hu 1998; Hu et al 1998, 2002, 2004; Kudritzki et
al 2000; Fynbo, Moller, \& Thomsen 2001; Pentericci et al 2000;
Stiavelli et al 2001; Ouchi et al 2001, 2003, 2004; Fujita et al 
2003; Shimasaku et al 2003, 2006; Kodaira et al 2003; Ajiki et 
al 2004; Taniguchi et al 2005; Venemans et al 2002, 2004).
However, the HST grism survey gives much broader redshift coverage
and higher spatial resolution over a smaller solid angle.  It 
yields immediate spectroscopic redshifts in many cases, and is
immune to the strong redshift selection effects introduced in
ground-based data by the forest of night sky OH emission lines.

After the data are reduced the spectra are extracted with
the software package aXe\footnote{
http://www.stecf.org/software/aXe}. 
We search for the emission lines on all extracted spectra
of the objects in the UDF field whose z-band (ACS F850LP filter)
magnitudes reach as faint as $z_{AB} = 29 \mag.$  While the
detection limit for continuum emission in the GRAPES
spectra is brighter than this ($z_{AB} \approx 27.2 \mag$;
see Pirzkal et al.~2004),
a fraction of such faint objects can have detectable emission lines.
The emission line search for this paper was performed on the 1D extracted
spectra, searching both in the spectra from individual PAs and on the 
combined spectra of all PAs.

To detect lines automatically, we wrote an IDL script (``emlinecull'')
that identifies and fits lines as follows.
We first remove the continuum by subtracting a median filtered version
of the spectrum from the original data.  We then determine whether the
resulting high-pass filtered spectrum shows evidence for emission
lines by sorting its pixels on signal to noise ratio and determining
the maximum {\it net} signal to noise ratio in a running total of (1,
2, ..., N) pixels.  (See Pirzkal et al. 2004 for further discussion of
this ``net significance'' parameter applied to unfiltered 1D spectra.)
In addition to ``net significance,'' we also calculate a cumulative
continuum flux (based on the median spectrum that we subtracted).  An
object is retained as a likely emission line source provided that (a)
its net significance exceeds $2.5$, and (b) the ratio of ``continuum''
(i.e., low-pass filtered flux) to ``line'' (i.e., high-pass filtered
flux) is $\le 10$.  
The second criterion eliminates broad features of low equivalent
width, thus avoiding a large catalog contamination by cool dwarf stars
with ``bumpy'' continuum spectra.  The precise line profile and other
details affect the precise correspondence between this cutoff and a
true equivalent width.  This criterion will of course introduce some
catalog incompleteness at low equivalent width, but this is inevitable
anyway, given the limited spectral resolution of the grism.

Once an object is identified as a likely emission line source, the
most significant peak in the filtered spectrum is identified, fitted
with a Gaussian profile, and subtracted.  This peak finding and
subtraction is iterated until no pixel remaining in the residual
spectrum exceeds the noise by a factor $> 2.5/\sqrt{2} = 1.77$.
(This implies a net line significance $\ga 2.5\sigma$ for any real
selected line, since the line spread function is never narrower
than two ACS pixels.)  The
output line list gives for each line the central wavelength, flux,
line width, continuum level, and equivalent width, along with
estimated uncertainties in each parameter.  The fitting code is based
on the IDL ``MPFIT'' package\footnote{
http://cow.physics.wisc.edu/$\sim$craigm/idl/idl.html} written by
Craig Markwardt.  

The fitting code then checks the measured significance of each line,
because line locations are initially based on the significance of
single pixels, while the final significance is based on the full
Gaussian fit to the line.  At this stage we discard any candidate line whose
final significance is $<1.8\sigma$ Because this $1.8\sigma$ cutoff is
applied in individual position angles, the final significance of a
line after combining data from multiple GRAPES position angles (see
below) will usually be $\ga 3$.  This step also discards any candidate lines
with a fitted FWHM $<10$\AA, because such narrow fitted widths are far
below the instrumental resolution and therefore indicate a noise spike
rather than a real spectral line.

The observed line width is usually determined by the angular size of
the object rather than its velocity width.  Formally it is a
convolution of the true emission line profile and the (projected)
spatial profile of the object multiplied by the $40$\AA\ per pixel
dispersion.  In general, this is dominated by the spatial term, unless
the line width exceeds $30,000 \kms \times (\theta / \hbox{arcsec})$,
where $\theta$ is the angular size of the object in the dispersion
direction.  Realistically, such
extreme linewidth to size ratios are only expected in broad-lined
active galactic nuclei (AGN). Here the observed spatial scale will be
the PSF size ($0.1''$) regardless of distance, and the minimum
velocity width for a resolved line becomes $3000 \kms$.

After the emission line lists from individual PAs
are generated through EmlineCull, we merge the lists according
to following criteria: 1) A line is retained
in one individual PA if the estimated contamination from 
overlapping spectra is less than 25\% of the (line plus
continuum) flux integrated over the line profile.
Otherwise, if the estimated contamination exceeds 25\%,
the line is simply discarded.
2) Two lines from two different PAs are considered
to be the same line if their wavelengths agree within their
combined line widths (FWHM), after possible offsets between
the cataloged object position and the location of strong
line emitting regions are taken into account.
3) A line is considered as a detection if it is detected 
in at least 2 PAs. After this merged emission line object list
is created, we visually examine all the individual spectra on 
both 1-dimensional extraction and 2-dimensional cut as a
sanity check.  This resulted in removal of some objects
from the list.  With the help of the emission line object
list generated from the PA combined spectra, we also visually
inspected all the combined spectra of the UDF objects down
to and sometimes fainter than $i_{AB} = 28.0 \mag$ as a completeness check.
This resulted in the addition of a few emission line objects.
Most of these were bright objects with broad emission lines of low
equivalent width, which were missed in the automated line finding
because they exceeded the threshold for continuum to line ratio.  

Among the parameters returned by ``emlinecull,'' the central
wavelength (and its associated error) are the most reliable.  The
other parameters are potentially affected by systematic effects.  Line
flux can be suppressed to some degree by the continuum subtraction
procedure, and is furthermore underestimated due to aperture losses in
the 1D extraction.  The equivalent width is vulnerable to sky
subtraction uncertainties for the faintest objects, and may also
suffer if the spatial distribution of line emitting regions does not
match that of continuum emitting regions, so that the equivalent width
within the extracted aperture is not the correct spatially averaged
equivalent width for the object.  As discussed above, the line width
is essentially a measure of object angular size for most objects.
We have therefore chosen to report only the line wavelengths and
approximate fluxes.

To improve the line flux estimates, we re-calculated them as follows.
First we fit the continuum to a pair of baseline regions, one on each 
side of the line, and subtract the resulting linear continuum fit 
from the spectrum.  The typical width for the continuum
fitting is about $2 \hbox{FWHM}$ on each side of the line.
We then do a direct integral of the the continuum-subtracted line to
determine its final flux estimate.
Because we do not know the spatial distribution of line emitting
regions in our sources {\it a priori}, we have chosen not to apply
aperture corrections to the line fluxes.  However, we can calculate
what the aperture correction would be if the line flux
were distributed like the continuum flux, using the direct
images from the HUDF.  To do so, we convolved the ACS i-band 
PSF with a gaussian of full width half maximum equal to the size of 
the object perpendicular to the grism direction, and then computed 
the fraction of the total flux falling within the grism extraction 
width.  The resulting aperture corrections were factors of 
$\sim 1.5$ to $\sim 3$, with a majority falling near the upper
end of this range.

The final emission line object list is presented in Table 1.  A total
of 115 objects are listed in Table 1, of which 101 objects have UDF ID
numbers from the released UDF catalogue.  Those without UDF ID are
either near a brighter object (with which they are blended in the UDF
catalog) or else lie just outside the UDF field. Note that the GRAPES
field of view covers a slightly larger area than that of UDF field due
to the combination of exposures at different orientations.  The total
number of lines listed in Table 1 is 147, as multiple emission lines
are detected in some galaxies.

The detected emission lines are then identified to 
the following template line list: a) Lyman-$\alpha$ ($\lambda$1216);
b) \oii; c) [O III] $\lambda$4959 and [O III] $\lambda$5007,
which are not resolved at the ACS grism resolution ($\sim 40$ \AA/pixel)
below redshift 1.0 or for extended sources, so are treated as a single feature
at \oiiic, their line ratio averaged wavelength;
and d) H$\alpha$ ($\lambda$6563).
These lines are chosen because they are usually the strongest
optical / near UV emission lines found in star forming galaxies. 
Some AGN emission lines are also used as a template in identifying AGNs.
But the AGNs are identified primarily based on their point-like morphologies
(Pirzkal et al. 2005).

The general line identification strategy is  summarized
as follows. 1) First we measure the photometric redshifts for
all the objects in the list using the photo-z code (Mobasher et al.~2004). 
Only 4 objects outside the UDF field of view  do not have photo-z 
measurements.  Images  from 7 filters are used  to determine the redshift.  
These are: 4 ACS bands (BViz), 2 NICMOS bands (JH) 
and 1 ISAAC band (K). 2) We then use the photometric 
redshifts as the input to identify observed emission lines to the template
lines and recalculate the redshift. If the recomputed redshift falls
within the 95\% confidence region of the photo-z redshift, we take
it as measured redshift.   3) Visually examine all the 
line identifications.  In last step, we have several different
approaches: a) If an object is found to have 2 or more lines,
we calculate the wavelength ratio of different lines to 
re-identify the lines and recalculate the redshift if necessary.
b) In both single and double emission line cases, if a relatively 
smooth break feature is found across the emission line region,
it is very likely that this feature is the  4000 \AA\ break,
and the corresponding line can be identified as \oii.
c) In the single emission line case, if a Lyman break feature is 
found near the emission line, and this Lyman break feature  
can be further confirmed with the broad band fluxes from the
direct images, then the corresponding
emission line is identified as Lyman-$\alpha$. 
d) If a line cannot be identified with
any method mentioned above, we simply present the measurements
of the line without deriving its redshift. In the final list, 
87\% of the sample has line redshifts that are consistent 
with their photometric redshifts.
Note that the GRAPES redshifts presented here do not in general 
provide an independent check on photometric redshift estimates,
since a photometric redshift is used to help identify the emission
lines in the GRAPES spectra.  An exception can be made for
objects with two GRAPES emission lines, in which case the wavelength
ratio of the lines is usually measured with sufficient precision to
identify them with no further information.

\begin{figure}
\plotone{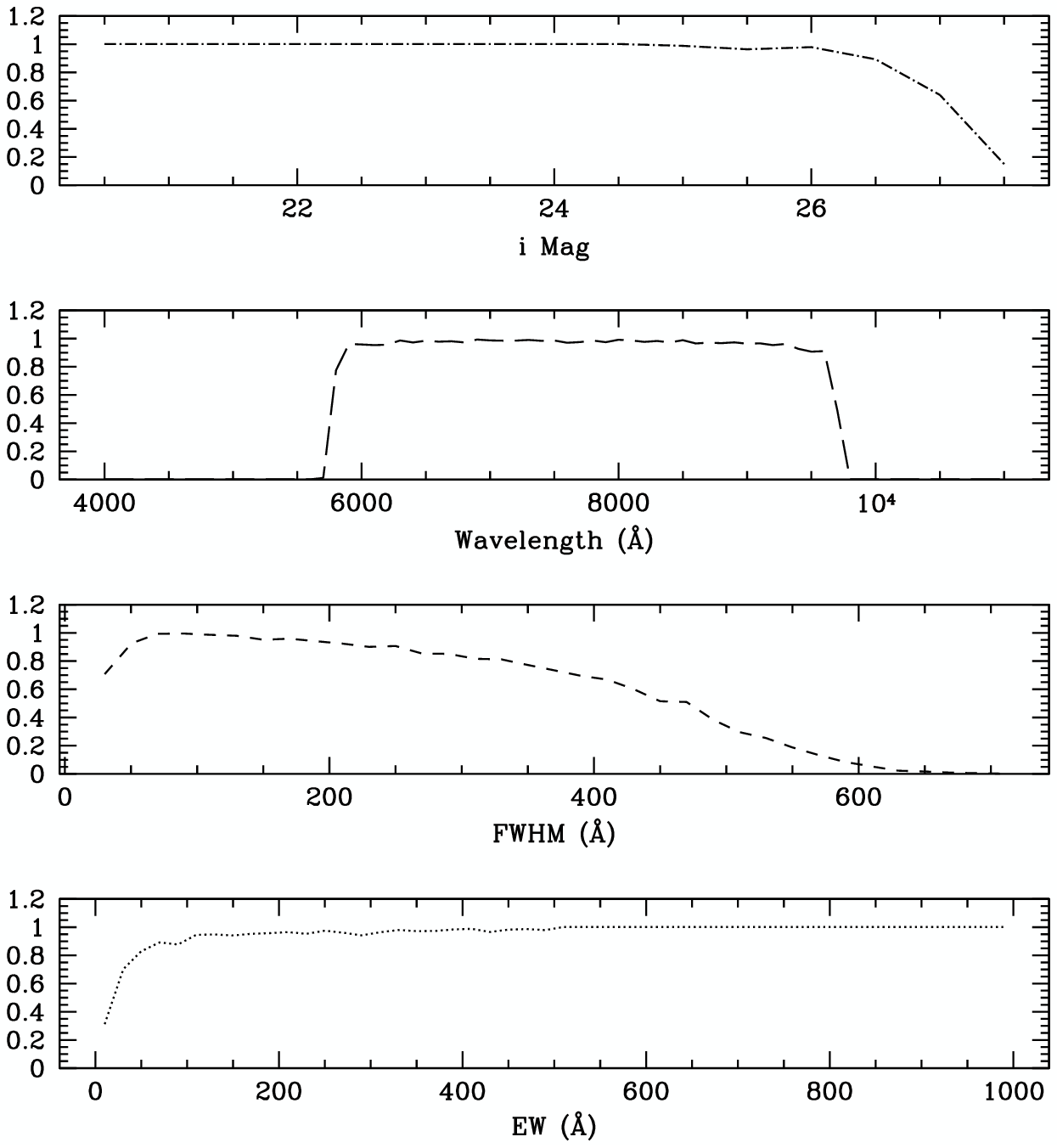}
\figcaption[f1]{The line recovery rate based on different
line width, equivalent width, central wavelength and host galaxy
brightness. In each panel, the parameter shown on the 
x axis is allowed to vary, while the remaining input parameters
are fixed at values that are generally favorable for line detections.
\label{f1}}
\end{figure}

To assess the selection effects in our sample, we performed extensive
Monte Carlo simulations of our line identification and measurement
procedures.  These simulations were performed in one dimension, using
Bruzual \& Charlot (2003) models for the continuum, with added
emission lines.  These lines were taken to be gaussians with various
widths and fluxes.  We selected 100~Myr, 5~Myr, and 1.4~Gyr templates
from the BC03 library, with metallicities of 0.08, 0.2 and 0.5 solar,
thus yielding 9 different templates with a range of 4000\AA\ break
strengths and stellar absorption line strengths.  In total, we
simulated 1.5 million emission lines, spanning a wide range of
continuum levels (e.g. scaled BC03 template), line fluxes, redshifts,
and object sizes (which determine the effective resolution of each
grism observation and the width of the observed lines).
For each simulation, we stored the line flux,
EW, i and z band broad band AB magnitude, redshift, and object size.
Each of the simulated spectrum was multiplied by the ACS grism
response function and resampled to the grism resolution of 40\AA\ per
pixel, prior to adding noise.  Our noise calculation included both
count noise from the input spectrum and the contribution of the sky
background ($\approx 0.1$ DN/pixel/s), and was scaled to match the
noise levels observed in the final, combined, GRAPES spectra.  We
identified spectral lines in each simulated spectrum using the same
``emlinecull'' script used to analyze the GRAPES data, and we
determined a simulated line to be successfully detected if its
measured line center was within one ACS grism resolution element
(i.e. 40\AA) of the simulated line center.  Using these simulations,
we were able to study the effect that wavelength, EW, line flux,
object size, and line blending (in the case of 4959/5007\AA) has on
the fraction of successfully detected lines.
Figure~\ref{f1} shows the recovery fraction in the simulations for
several input parameters.  In each case, the remaining parameters
are fixed at values that allow for easy line identification.
% To a first approximation, the overall selection function for a
% given situation can be estimated as a product of the selection
% probabilities for the input magnitude, equivalent width,
% line wavelength, and object size.

\lya\ presents a special case here because it is invariably located
atop a significant spectral break when observed at the relevant
redshifts.  This means that the line not only needs to be significantly
detected with respect to the photon noise  ($s/n > 2.5$ as usual),
but also must pass an equivalent width threshold that depends on the
spectral resolution, in order to appear as a distinct emission line that
the algorithm will identify. 
We characterized this effect 
by convolving a model spectrum of a \lya\ emission line
atop the corresponding \lya\ break with line spread functions
of width $60 \hbox{\AA} \la \Delta \lambda \la 250$\AA, 
adding random noise, and running our detection algorithm.  
In the noise-free case, $\hbox{EW} \ga 0.4 \Delta \lambda$ is 
required for the line to be recovered.  If we lower the continuum
signal-to-noise ratio, the EW rises.  The required equivalent
widths for 80\%  completeness, given $s/n \approx 100$, $10$,
$4.5$, $2$, and $1$ per pixel in the continuum,
become $\approx 0.5 \Delta \lambda$, $ 0.75 \Delta \lambda$, 
$\Delta \lambda$, $1.5 \Delta \lambda$, and
$3 \Delta \lambda$, respectively.  In the limit of vanishingly
small continuum, the problem reverts to detecting an isolated line, 
and for $s/n < 1$ per pixel,
the threshold equivalent width scales as $1/(s/n)$ as expected.

\section{Results}
Table 1 lists a total of 115 galaxies that have detectable emission lines.
Among these, 9 are high redshift Lyman-$\alpha$ emitters; at least
3 (and possibly up to 6) are active galactic nuclei (AGN);
and the remainder are star forming galaxies detected in some
combination of their \oii, \oiiipair, and
H$\alpha$ lines. In this latter category, 11 of them are 
detected with both \oii\ and \oiiipair\, and 16 are detected with both
\oiiipair\ and H$_{\alpha}$.
The first column of this table lists the UDF ID of the objects.
An ID of ``$\le -100$'' means that the object is outside or near the edge 
of the UDF field so it does not have a UDF ID. An ID between 0 and
-100 means that the object is too close to an extended galaxy so it 
might be taken as part of that galaxy thus not assigned a UDF ID.
The objects added into the 
line list after visual inspection of their spectra are marked with \dag.
Column ``i-mag'' is the i-band AB magnitude of the object. 
Column ``wavelength'' is the central wavelength of the detected line,
along with its $1\sigma$ statistical error estimate.
Column ``Line Flux'' is line flux in the observed frame, as measured
directly from the 1D spectra {\it without aperture correction}.  The
true line flux is in general larger by a factor of $1.5$ to $3$.
Column ``Redshift'' is redshift determined from the emission
lines, with a redshift of ``-1'' indicating that this line is not 
identified.
Column ``Line'' is the physical identification of the observed
line.
Some examples of the emission line galaxy spectra are presented
in figure~\ref{f3}.

Figure~\ref{f4}  shows the redshift distributions of the
emission line galaxies according to the detected lines.  The redshift
range covered by each emission line is effectively set by the
wavelength coverage of the grism and the
rest wavelengths of the lines.  The grism response extends 
from 5500\AA\ to 10500\AA.
In practice, we find we can detect emission lines usefully over
the wavelength range 5700 \AA\ $\la \lambda \la$ 9700\AA,
which corresponds to a grism throughput $\ga 25\%$ of the
peak throughput.  In this figure, a curve of
the number (per redshift bin) of emission line galaxies that would be expected
in the corresponding redshift (bin) are also overplotted. 
To derive these curves, we first generated a modified line
luminosity function directly from the GRAPES data, 
using the $1/V$ method and the empirical sample, but disregarding
both selection effects and evolution.  We then used this function 
to estimate the number of objects that ought to have been detected
in each redshift bin.
The minimum luminosity for line detection was empirically calibrated
to the faintest detected GRAPES emission lines, with the redshift
dependence based on the grism sensitivity and the luminosity distance
calculated in the current concordance cosmology 
($\Omega_m = 0.3, \Omega_\Lambda = 0.7, H_0 = 70 \kmsMpc$; 
see Spergel et al. 2003, 2006).  The dotted curves are 
in some way an interpolation of the observed number-redshift
counts, but they improve on a direct interpolation by
incorporating our knowledge of the cosmology and instrument
properties.  Because the histograms are plotted based on detected 
lines, galaxies can 
contribute to more than one histogram if they have multiple
detected emission lines.

\section{Discussion}
The GRAPES spectroscopy goes to much fainter continuum flux levels
than is typical for ground-based followup spectroscopy.  The
median broad band magnitude for the emission line objects in table~1
is $i_{AB}=24.67 \mag$ (see also figure~\ref{f2}).  
Over 40\% of the sample has $i_{AB}>25 \mag$, which is
about the faintest magnitude level routinely targeted for ground-based
followup spectroscopy, and 15\% is fainter than $i_{AB}=26 \mag$.  It is
thus likely that 30--40\% of our emission line sample would be missed
by ground-based followup efforts.  The GRAPES sample thus provides a
unique resource for studying emission line luminosity functions and
the star formation rate density (SFRD) at moderate redshifts. 
The results of such studies will be presented in the 
companion paper (Gronwall et al. 2007). 
The morphologies of the emission line galaxies are presented in 
Pirzkal et al. (2006). 

The 115 redshifts from this emission line catalog represent a
fraction $\sim 5\%$ of the sources in the Hubble Ultra Deep Field
brighter than 28th magnitude (AB).  The primary factors leading 
to this comparatively small fraction are the selection effects
demonstrated in figure~1.  For an object to be selected in our
emission line catalog with a reasonable probability, it must
have $i_{AB} \la 26.5$, $\hbox{EW} \ga 50$\AA, size $< 0.75''$,
and an emission line with observed wavelength $5800{\rm \AA} \la
\lambda \la 9600{\rm \AA}$.  Indeed, if we only consider objects
with $i_{AB} < 26.5$, the redshift completeness rises to $9.5\%$.
The GRAPES survey has also published continuum break redshifts of 
Lyman break galaxies (Malhotra et al 2005) and distant elliptical
galaxies (Daddi et al 2005; Pasquali et al 2005), and there are
additional redshift samples in progress for later type galaxies
with $4000$\AA\ breaks (Hathi et al 2007, in preparation) and
for an overarching sample of galaxies with significant spectroscopic
continuum information in the GRAPES spectra (Ryan et al 
2007, see below).  So the spectroscopic success rate is reasonably
high for sources where sufficient information can be expected
given the properties of the instrument and data set.  In particular,
the fraction of redshifts lost to crowding and overlap is
modest, because of our multiple roll angle observing strategy.

As a sanity check, we compared our measured redshifts with those
available in the Chandra Deep Field South and found good agreement.
We use redshifts from five available references (Vanzella
et al.~2005; Le Fevre et al.~2005, Szokoly et al.~2004, 
Vanzella et al.~2006, and Grazian et al.~2006).
The results of this comparison are given in table~2.
Both the overall agreement and the 
agreement with individual references was generally good.
For example, we found that there are 7 objects in common
in our emission line catalog and the VLT/FORS2 
catalog of Vanzella et al.~2005.  Among these, redshifts
are in agreement for 6 objects.  Moreover, the one 
that does not agree, UDF -100, also shows up in Le Fevre et al.~2004,
where its redshift {\it does} agree with our measurement.
Among the 9 objects in Le Fevre, only one (UDF3484) disagrees.
We also found  4 out of 5 in agreement with Szokoly et al.~2004
(here the object that does not agree is UDF4445, a 2-line object).
The total sample is 23 objects with both GRAPES and ground-based
redshifts available.  Among these, $2.5$ show ``catastrophic'' redshift
mismatches (where the ``0.5'' object is our ID -100, 
with two inconsistent ground-based redshifts).  This 11\% failure rate
is essentially the same as the catastrophic failure rate for photometric 
redshifts, and for essentially the same reason:  Most of our redshifts
are constrained to match a photometric redshift at the outset.
The catastrophic failures in our list correspond to mis-identified
emission lines.

In a future paper (Ryan et al 2007) we will combine the 
continuum shape information from the grism with multi-band optical
and near-IR photometry to derive spectro-photometric redshifts for
the GRAPES data set.  This effort may reduce the catastrophic
failure rate in line identifications, since the extra information
from the grism continuum will help distinguish among the redshift
likelihood peaks in photometric redshift fitting.

To estimate our redshift accuracy, we study the cleanest subset
of these object with ground-based and GRAPES redshifts.
We first exclude the catastrophic failures from consideration (excluding
ID -100 along with the other two).  Our catalog
also has a few lines from the manually identified emission line
galaxies where we lack an automated line wavelength uncertainty
estimate, and we exclude these also for the present.
Among the remaining 15 overlap objects, the RMS redshift difference
between GRAPES and ground-based data is $\hbox{RMS}(\Delta z) = 0.008$.
For comparison, the median estimated redshift uncertainty from
our line centroiding algorithm is $\hbox{median}(\delta \lambda /
\lambda_{rest}) = 0.009$.
Thus, we infer a typical redshift uncertainty just below $0.01$
for our emission line catalog.
Examining the redshift residuals $z_{GRAPES} - z_{VLT}$ 
as a function of redshift shows weak evidence for a systematic
offset at $z \la 0.3$, with $\langle z_{GRAPES} - z_{VLT} \rangle
\sim 0.01$, but this is based on only 3 or 4 data points and 
may be a coincidence.

If we examine our subsample with two emission lines and compare the
redshifts derived separately from the two lines, we reach a similar
conclusion-- the offset between these measurements is again fully
consistent with characteristic redshift errors $\delta z \la 0.01$.

The comparison with ground-based spectra also allows us to 
estimate the systematic error floor in our line wavelength measurements:
A systematic error component of $\sim 12$\AA\ is quite sufficient
to account for the observed redshift offset between
ground and GRAPES data for those lines whose formal wavelength
uncertainty $\delta \lambda < 5$\AA.  A larger comparison 
sample might refine this estimate, but is unlikely to change
it dramatically.

\section{Conclusion}

We find that a deep spectroscopic survey
like GRAPES offers a unique opportunity to identify emission lines
and determine redshifts for faint galaxies. 
In this paper, we present a catalog of emission lines
identified in GRAPES, including wavelength measurements, flux
estimates, line identifications, and redshifts.  
Over 40\% of the sample comes from objects fainter than 
the typical continuum magnitude limit for ground-based multiobject 
spectroscopic followup programs.  These
objects might never have been identified as emission line galaxies
without a space based observation as ours.  Based on comparison
with ground-based spectra for a subset of our objects, we infer
a typical redshift accuracy of $\delta z = 0.009$ for our
catalog.

\acknowledgments
This work was supported by grant GO-09793.01-A from the Space 
Telescope Science Institute, which is operated by AURA under 
NASA contract NAS5-26555. ED acknowledge support from 
NASA through the Spitzer Fellowship Program, under award 1268429.
This project has made use of the aXe 
extraction software, produced by ST-ECF, Garching, Germany.
We also made use of the "mpfit" IDL library, and we thank Craig 
Markwardt for making this package public.

\begin{figure}
\plotone{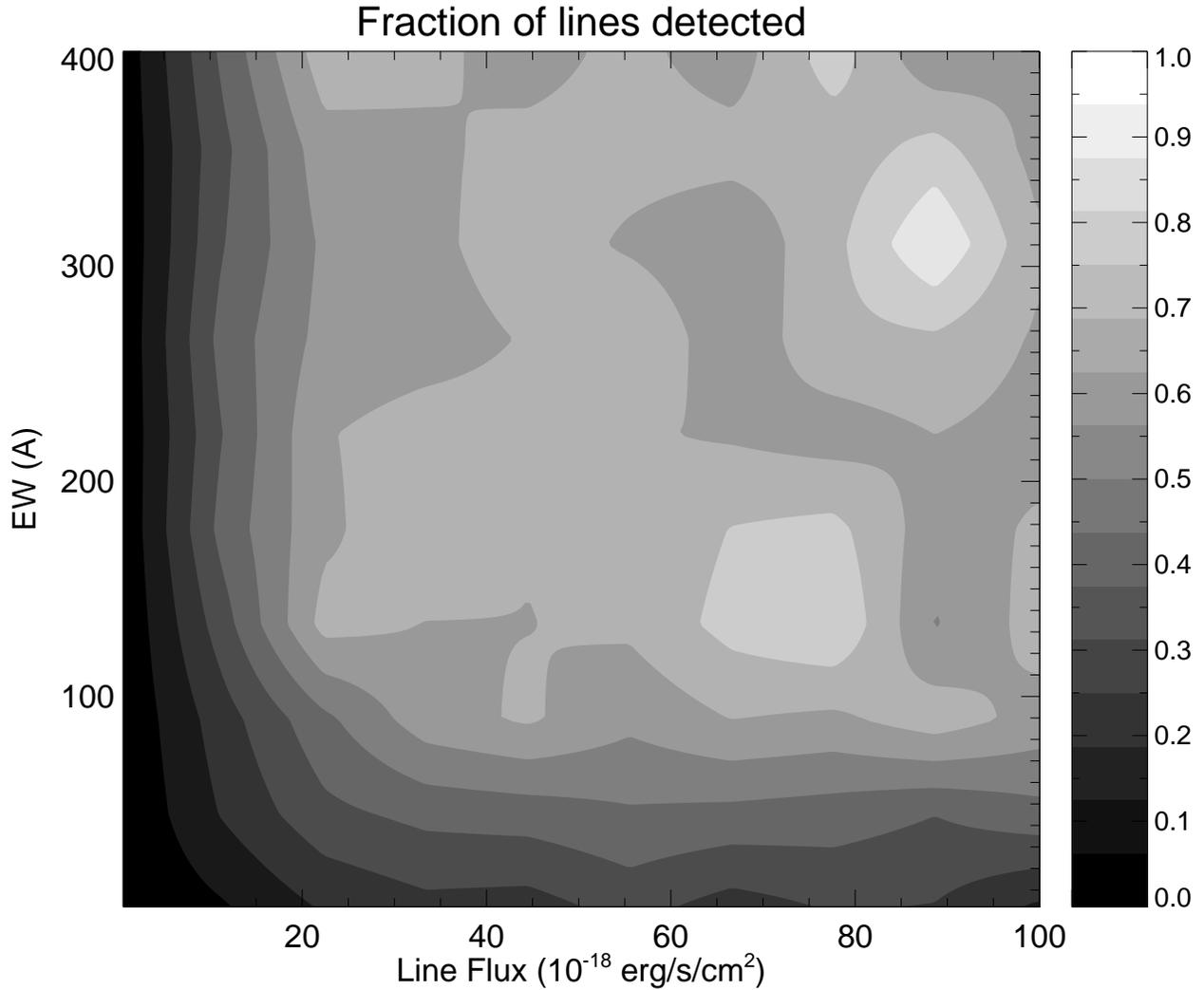} \figcaption[f1b]{The line recovery rate as a
two-dimensional function of line flux and equivalent width.  In this
figure (unlike figure~1), the other relevant parameters are allowed
to vary over a wide range.  Thus, the ``plateau'' region in the
contour plot with selection probability $\ga 60\%$ corresponds to
the region where the EW and line flux are sufficient for selection.
Those lines {\it not} recovered in this portion of the plot were
missed due to an unfavorable choice of line width or
wavelength. \label{f1b}}
\end{figure}

\begin{figure}
\plotone{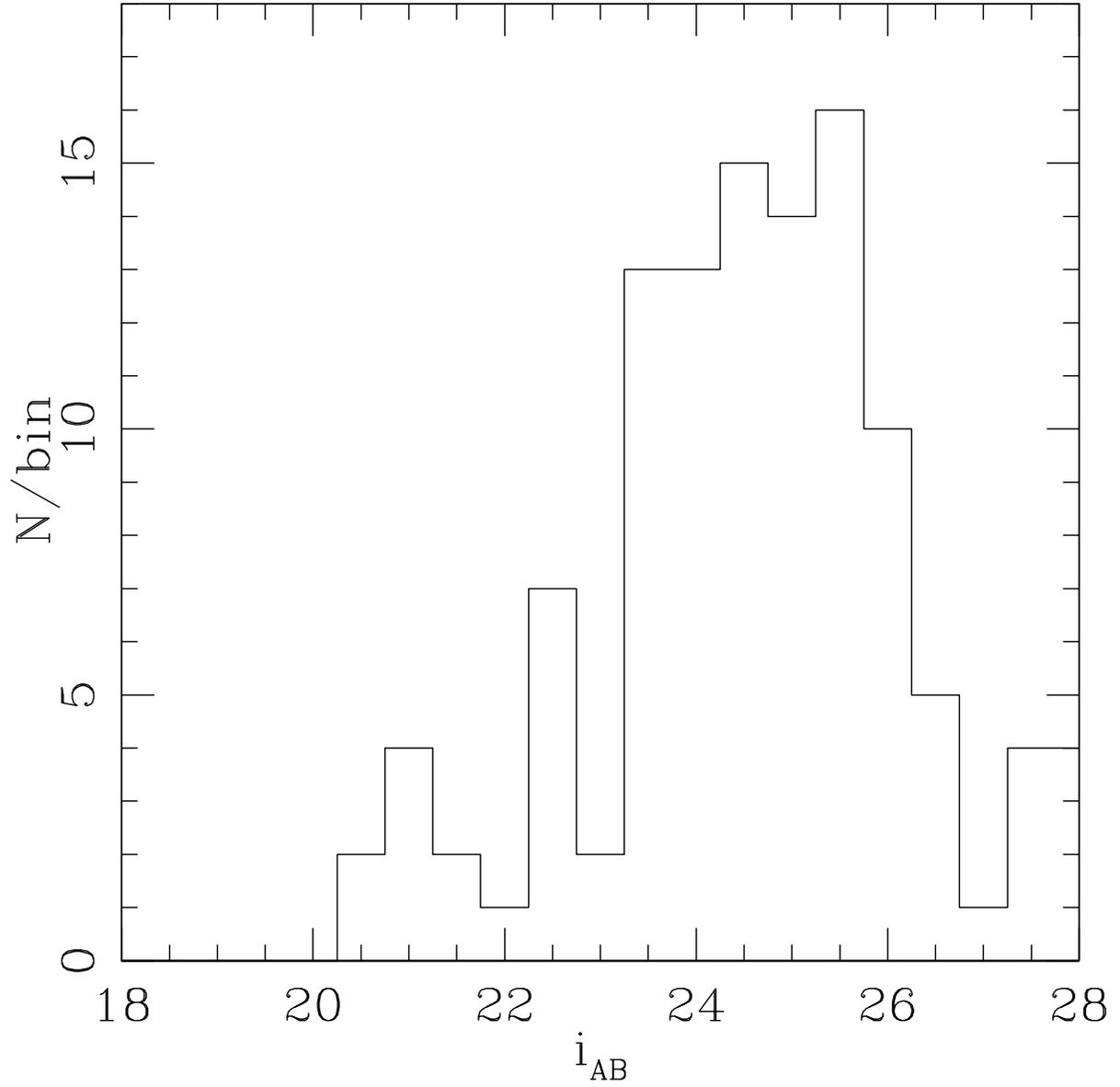}
\figcaption[f2]{The histogram of i-band magnitude of
all the emission line objects listed in Table 1. 
The binsize of the histogram is 0.5 magnitude.
\label{f2}}
\end{figure}

\clearpage
\begin{figure}
\plotone{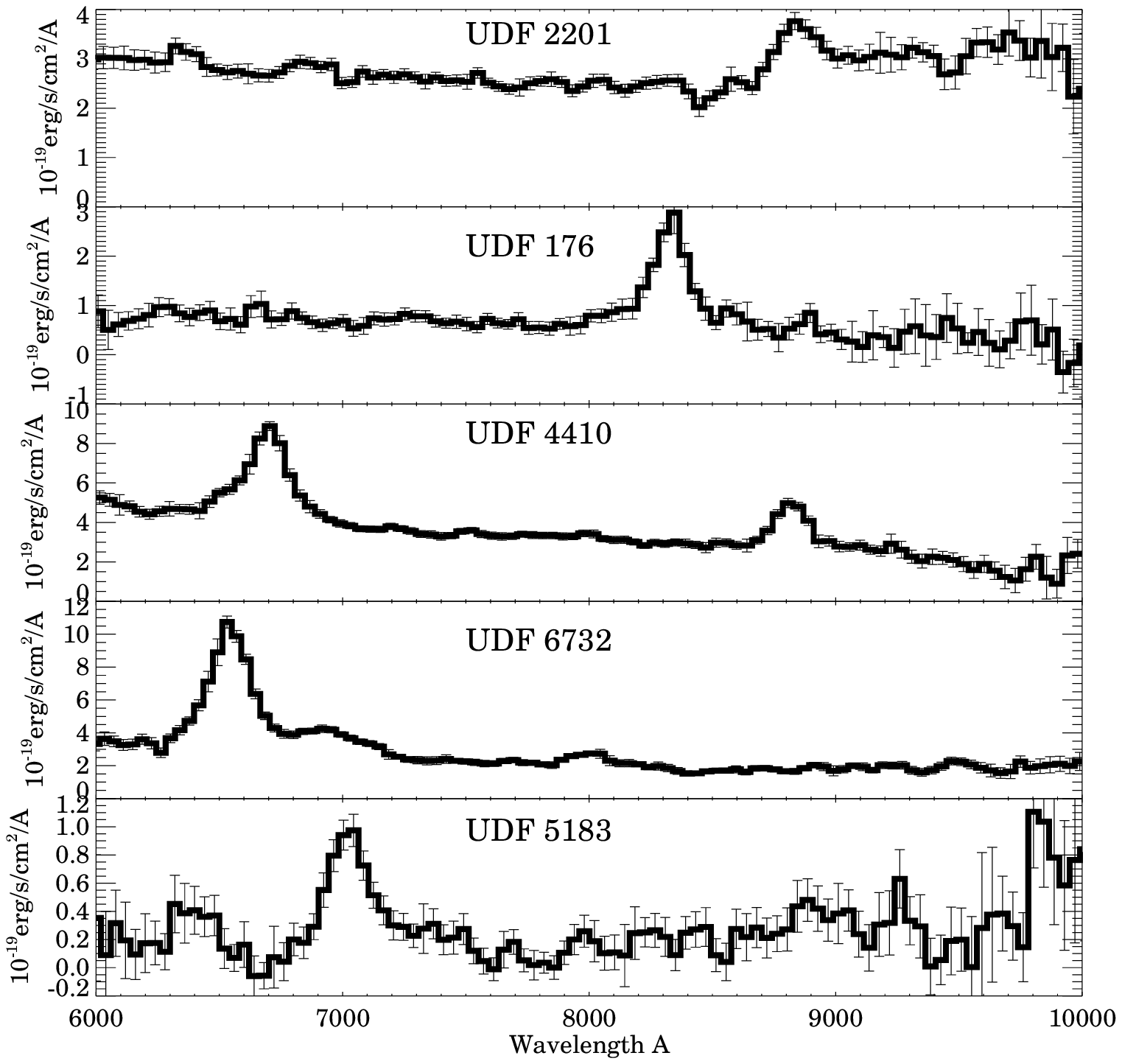}
\figcaption[f3]{Some examples of the ACS grism spectra 
from the emission line galaxy catalog. From the top to the bottom 
panels are: UDF 2201, with \oii\ at $z=1.377$; UDF 176, with \oiiipair 
at $z=0.667$; UDF 4410, with \oiiipair\ and $H\alpha$ at $z=0.346$;
UDF 6732, a quasar with \civ\ and \ciii\ at $z=3.19$; 
and UDF 5183, a Ly$\alpha$ emitter at $z=4.78$.
The apparent continuum level in UDF~5183 (at magnitude $i_{AB} \approx
27.4 \mag$) should be treated with caution, as it is below the systematic 
error level in GRAPES sky subtraction (a few$ \times 10^{-4}$
 counts/sec/pixel; see Pirzkal et al 2004).
\label{f3}}
\end{figure}

\begin{figure}
\plotone{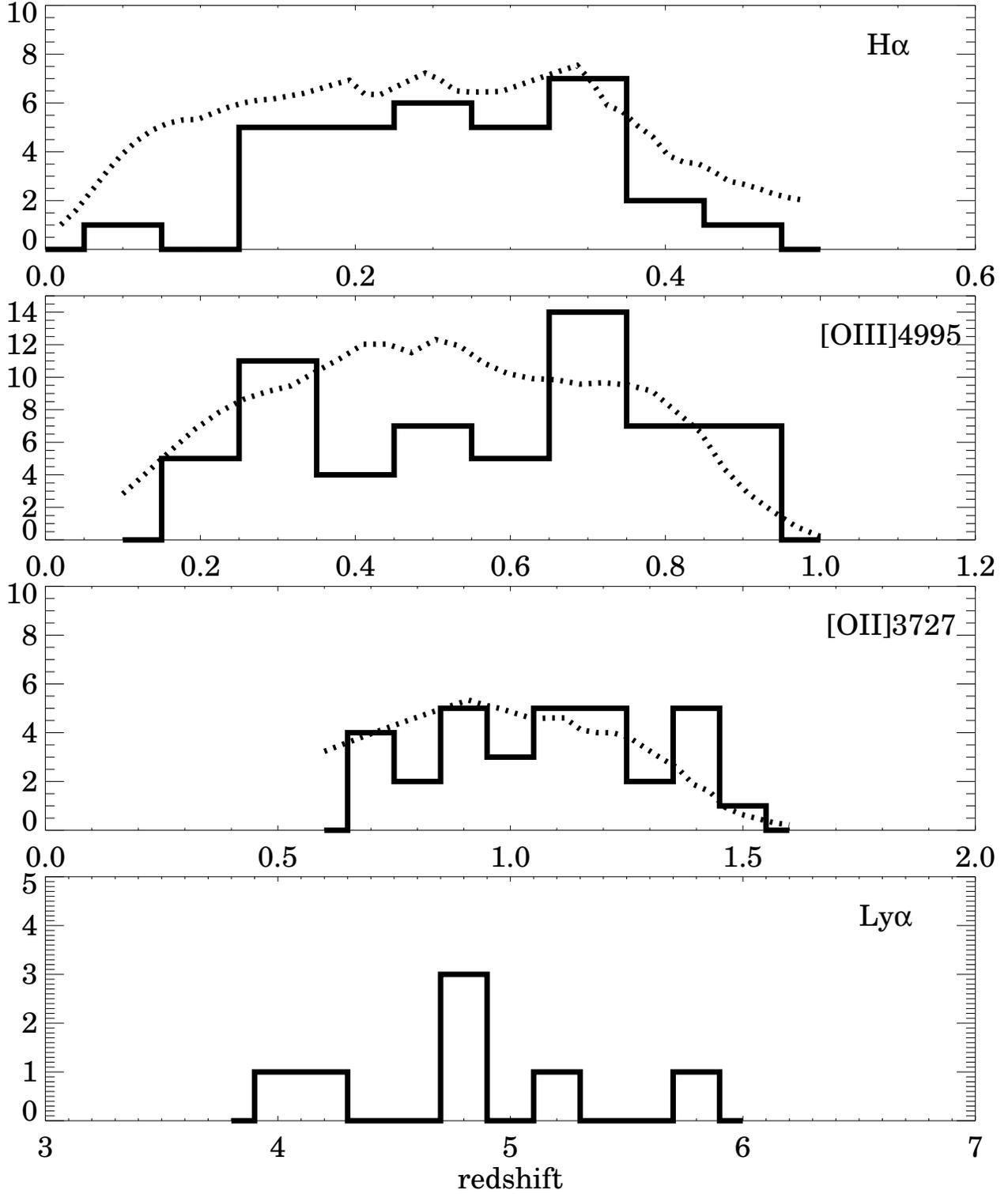}
\figcaption[f4]{\scriptsize{The distribution of the redshifts from different lines.
From top panel to the bottom are $H\alpha$, \oiiipair, 
\oii\ and Ly$\alpha$. Their binsize in redshift z are 0.05,
0.1, 0.1 and 0.2 respectively. 
Each histogram is overplotted with the number of objects that
would be expected in that bin based on the minimum line luminosity
required for detection as a function of redshift and rest wavelength,
and on line luminosity functions derived directly from GRAPES data.
Comparing the observed redshift histogram with this curve
allows us to identify overdense and underdense regions along
the line of sight.  
While individual peaks are not strongly significant given our
sample size, we note that the peak at $z\approx 0.7$ in the \oiii\ redshift
distribution corresponds to two closely spaced peaks that
have been previously noted in the CDF-South field at $z\approx 0.67$
and $z\approx 0.73$ (Vanzella et al 2005).
\label{f4}}}
\end{figure}

\clearpage
\begin{deluxetable}{cccccccc}
\rotate
\tabletypesize{\small}
\tablewidth{0pc}
\tablecaption{Emission Line Objects in UDF Field\label{common}}
\tablehead{ 
\colhead{UDF}&\colhead{RA}&\colhead{DEC}&\colhead{i-mag}&
\colhead{Wavelength\tablenotemark{a}}
&\colhead{Measured Line Flux\tablenotemark{b}}&
\colhead{Redshift}&\colhead{Line} \\
\colhead{ID}&\colhead{}&\colhead{}&\colhead{(AB)}&
\colhead{\AA}&\colhead{$10^{-18} erg/s/cm^2$}&
\colhead{}&\colhead{}
}

\startdata
    -100$^\dag$& 53.1773491&-27.7639294&22.439&  7995.7 $\pm$   6.7&   105&  0.601& [OIII]$\lambda\lambda$4959,5007 \\
           -101& 53.1878090&-27.7726192&22.681&  7470.4 $\pm$  37.3&   107&  0.138&           H$\alpha$ \\
           -102& 53.2072487&-27.7847309&23.884&  7617.9 $\pm$  42.3&    18&  0.161&           H$\alpha$ \\
           -103& 53.1970100&-27.7805996&24.127&  6485.5 $\pm$  32.2&    24&  0.740&  [OII]$\lambda$3727 \\
               &           &           &      &  8721.9 $\pm$  42.1&    18&  0.746& [OIII]$\lambda\lambda$4959,5007 \\
    -104$^\dag$& 53.1931000&-27.7755508&25.233&  8265.5 $\pm$   1.4&    13&  1.218&  [OII]$\lambda$3727 \\
           -105& 53.1855392&-27.8111000&24.417&  6663.6 $\pm$  28.0&     7&  0.334& [OIII]$\lambda\lambda$4959,5007 \\
               &           &           &      &  8748.8 $\pm$  32.2&    17&  0.333&           H$\alpha$ \\
           -106& 53.1649399&-27.7558804&25.263&  7535.7 $\pm$  41.3&    19& -1.000&                     \\
    -107$^\dag$& 53.1795387&-27.7661705&25.057&  8336.5 $\pm$  22.0&    34&  0.669& [OIII]$\lambda\lambda$4959,5007 \\
    -108$^\dag$& 53.1948090&-27.7733498&25.576&  8063.1 $\pm$   6.5&    27&  0.614& [OIII]$\lambda\lambda$4959,5007 \\
    -109$^\dag$& 53.1554108&-27.8261108&25.431&  8009.8 $\pm$  11.2&    37&  0.604& [OIII]$\lambda\lambda$4959,5007 \\
           -110& 53.1681290&-27.7565002&25.536&  7867.2 $\pm$  44.0&    18& -1.000&                     \\
             -1& 53.1843987&-27.8051605&25.383&  6253.9 $\pm$  41.2&    11&  0.678&  [OII]$\lambda$3727 \\
               &           &           &      &  8347.4 $\pm$  29.9&    20&  0.671& [OIII]$\lambda\lambda$4959,5007 \\
             33& 53.1632500&-27.8267040&25.591&  9201.5 $\pm$  50.5&    14&  0.842& [OIII]$\lambda\lambda$4959,5007 \\
            119& 53.1660042&-27.8238735&27.364&  7388.9 $\pm$  47.9&     7&  0.479& [OIII]$\lambda\lambda$4959,5007 \\
            166& 53.1605415&-27.8226891&25.650&  8104.6 $\pm$  38.7&     6&  1.897& [MgII]$\lambda$2798 \\
            176& 53.1583786&-27.8228283&25.728&  8327.0 $\pm$  39.4&    27&  0.667& [OIII]$\lambda\lambda$4959,5007 \\
            178& 53.1610794&-27.8219051&23.753&  8313.5 $\pm$  43.3&    27&  0.664& [OIII]$\lambda\lambda$4959,5007 \\
            206& 53.1717300&-27.8217850&23.503&  7181.3 $\pm$  25.4&     9&  0.927&  [OII]$\lambda$3727 \\
               &           &           &      &  9642.8 $\pm$  27.3&    76&  0.930& [OIII]$\lambda\lambda$4959,5007 \\
            631& 53.1670227&-27.8169975&26.463&  6234.5 $\pm$  17.9&    20&  4.127&          Ly$\alpha$ \\
            648& 53.1607285&-27.8162994&24.754&  7492.2 $\pm$  32.0&    16&  0.142&           H$\alpha$ \\
            703& 53.1734314&-27.8157158&24.618&  8547.7 $\pm$  30.0&    20&  0.711& [OIII]$\lambda\lambda$4959,5007 \\
            704& 53.1788826&-27.8159561&25.221&  8692.8 $\pm$  36.6&    11&  0.740& [OIII]$\lambda\lambda$4959,5007 \\
            712& 53.1783562&-27.8162518&27.296&  7537.0 $\pm$  49.1&     6&  5.198&          Ly$\alpha$ \\
     900$^\dag$& 53.1851387&-27.8052711&20.354&  9313.6 $\pm$   4.3&   280&  0.419&           H$\alpha$ \\
           1000& 53.1531181&-27.8120880&23.301&  6135.2 $\pm$  43.8&    39&  0.228& [OIII]$\lambda\lambda$4959,5007 \\
               &           &           &      &  8083.0 $\pm$  51.2&    56&  0.231&           H$\alpha$ \\
           1016& 53.1813507&-27.8126469&25.761&  6860.1 $\pm$  38.9&     6&  0.841&  [OII]$\lambda$3727 \\
               &           &           &      &  9155.1 $\pm$  30.6&    32&  0.833& [OIII]$\lambda\lambda$4959,5007 \\
    1309$^\dag$& 53.1512070&-27.8095436&24.600&  8296.3 $\pm$  17.2&    23& -1.000&                     \\
           1322& 53.1506081&-27.8094997&24.624&  6563.8 $\pm$  52.6&     6&  0.314& [OIII]$\lambda\lambda$4959,5007 \\
               &           &           &      &  8606.7 $\pm$  21.7&    33&  0.311&           H$\alpha$ \\
           1344& 53.1536598&-27.8089428&23.835&  6805.2 $\pm$  59.6&    13&  0.037&           H$\alpha$ \\
           1347& 53.1510468&-27.8094597&24.665&  7753.5 $\pm$  32.2&     5&  1.080&  [OII]$\lambda$3727 \\
           1446& 53.1632233&-27.8089867&26.509&  9226.4 $\pm$  44.6&    10&  2.470&                     \\
           1752& 53.1705093&-27.8066654&25.246&  8341.0 $\pm$  25.6&    10&  1.238&  [OII]$\lambda$3727 \\
           1889& 53.1732178&-27.8061523&25.852&  9063.5 $\pm$  26.5&    17&  1.432&  [OII]$\lambda$3727 \\
           2088& 53.1540489&-27.8051834&26.694&  8329.9 $\pm$  39.7&     9&  0.668& [OIII]$\lambda\lambda$4959,5007 \\
           2162& 53.1487541&-27.8043003&24.029&  7135.4 $\pm$  49.6&    23&  0.914&  [OII]$\lambda$3727 \\
               &           &           &      &  9578.8 $\pm$  40.2&    60&  0.918& [OIII]$\lambda\lambda$4959,5007 \\
           2201& 53.1416206&-27.8040810&24.428&  8859.2 $\pm$  43.4&    15&  1.377&  [OII]$\lambda$3727 \\
           2241& 53.1377258&-27.8042164&25.860&  7674.1 $\pm$  40.6&    26&  0.536& [OIII]$\lambda\lambda$4959,5007 \\
           2848& 53.1472511&-27.8008347&26.087&  6378.2 $\pm$  28.0&    26&  0.277& [OIII]$\lambda\lambda$4959,5007 \\
               &           &           &      &  8385.9 $\pm$  25.0&    11&  0.278&           H$\alpha$ \\
           2927& 53.1478958&-27.7996979&23.945&  7274.1 $\pm$  44.5&     7&  0.456& [OIII]$\lambda\lambda$4959,5007 \\
           2974& 53.1733017&-27.7992401&23.668&  7564.0 $\pm$  28.7&    24&  0.153&           H$\alpha$ \\
               &           &           &      &  9569.9 $\pm$  56.0&    37& -1.000&                     \\
           2998& 53.1512070&-27.7987099&23.647&  8023.2 $\pm$  49.3&    38&  0.222&           H$\alpha$ \\
           3068& 53.1781006&-27.7999439&25.940&  8461.9 $\pm$  38.3&    13&  1.270&  [OII]$\lambda$3727 \\
           3484& 53.1980057&-27.7967453&23.088&  7017.6 $\pm$  72.1&    25&  0.405& [OIII]$\lambda\lambda$4959,5007 \\
           3823& 53.1613274&-27.7958012&24.431&  9266.9 $\pm$  32.9&    15&  1.486&  [OII]$\lambda$3727 \\
           3869& 53.1880836&-27.7957401&24.232&  7063.7 $\pm$  41.5&    13&  0.414& [OIII]$\lambda\lambda$4959,5007 \\
           3914& 53.1873589&-27.7942257&22.549&  6660.7 $\pm$  28.3&   256&  0.333& [OIII]$\lambda\lambda$4959,5007 \\
               &           &           &      &  8815.1 $\pm$  32.8&   166&  0.343&           H$\alpha$ \\
           3959& 53.1794853&-27.7960243&28.023&  9203.7 $\pm$  45.5&     4&  0.843& [OIII]$\lambda\lambda$4959,5007 \\
           3977& 53.1558685&-27.7949009&23.462&  7833.1 $\pm$  39.4&    17&  1.102&  [OII]$\lambda$3727 \\
           4029& 53.1663704&-27.7956104&26.729&  7688.9 $\pm$  30.6&    10&  0.539& [OIII]$\lambda\lambda$4959,5007 \\
           4084& 53.1278381&-27.7947979&24.050&  7868.7 $\pm$  32.3&     5&  1.111&  [OII]$\lambda$3727 \\
           4120& 53.1839104&-27.7954140&27.876&  8661.0 $\pm$  32.6&    19&  2.095& [MgII]$\lambda$2798 \\
           4126& 53.1541138&-27.7950897&25.988&  8622.6 $\pm$  41.1&    12&  0.726& [OIII]$\lambda\lambda$4959,5007 \\
    4142$^\dag$& 53.1841469&-27.7926502&21.628&  6437.1 $\pm$   0.0&    51&  0.727&  [OII]$\lambda$3727 \\
               &           &           &      &  8635.1 $\pm$   0.0&   148&  0.729& [OIII]$\lambda\lambda$4959,5007 \\
           4172& 53.1255722&-27.7949314&25.835&  8789.3 $\pm$  38.5&     8&  1.358&  [OII]$\lambda$3727 \\
           4262& 53.1619377&-27.7925472&22.748&  7316.5 $\pm$  41.1&    61&  0.465& [OIII]$\lambda\lambda$4959,5007 \\
           4315& 53.1546211&-27.7932377&22.668&  6150.1 $\pm$  11.1&    11&  0.231& [OIII]$\lambda\lambda$4959,5007 \\
               &           &           &      &  8081.0 $\pm$   7.8&    20&  0.231&           H$\alpha$ \\
           4371& 53.1822014&-27.7939930&24.954&  8614.0 $\pm$  39.2&    13&  0.725& [OIII]$\lambda\lambda$4959,5007 \\
           4396& 53.1491623&-27.7929745&24.445&  8266.3 $\pm$  40.0&    17&  1.218&  [OII]$\lambda$3727 \\
           4410& 53.2002335&-27.7937489&24.158&  6720.9 $\pm$  38.9&    92&  0.345& [OIII]$\lambda\lambda$4959,5007 \\
               &           &           &      &  8848.0 $\pm$  33.8&    40&  0.348&           H$\alpha$ \\
           4442& 53.1640625&-27.7942142&29.564&  8221.0 $\pm$  29.4&     7&  5.761&          Ly$\alpha$ \\
           4445& 53.1615601&-27.7922573&21.130&  7178.4 $\pm$  42.4&   137&  0.926&  [OII]$\lambda$3727 \\
               &           &           &      &  9599.8 $\pm$  17.5&   705&  0.922& [OIII]$\lambda\lambda$4959,5007 \\
           4496& 53.1530037&-27.7936897&26.546&  8599.2 $\pm$  74.1&    19&  1.307&  [OII]$\lambda$3727 \\
           4816& 53.1840210&-27.7915077&24.560&  8311.0 $\pm$  33.0&    12&  1.230&  [OII]$\lambda$3727 \\
           4825& 53.1673279&-27.7918644&23.961&  6698.5 $\pm$  14.6&    14&  0.341& [OIII]$\lambda\lambda$4959,5007 \\
               &           &           &      &  8781.7 $\pm$  22.8&     7&  0.338&           H$\alpha$ \\
    4929$^\dag$& 53.1879463&-27.7900028&20.792&  9436.0 $\pm$   0.0&   590&  0.438&           H$\alpha$ \\
           4981& 53.1444244&-27.7911167&24.692&  9056.2 $\pm$  28.1&     6&  1.430&  [OII]$\lambda$3727 \\
           5183& 53.1437798&-27.7908649&27.417&  7032.9 $\pm$  39.8&    11&  4.784&          Ly$\alpha$ \\
           5187& 53.1848183&-27.7899323&23.702&  7313.1 $\pm$  27.0&    14&  0.962&  [OII]$\lambda$3727 \\
           5225\tablenotemark{c}& 53.1385743&-27.7902115&25.953&  7933.2 $\pm$  33.9&    22 & 5.42 &          Ly$\alpha$           \\
           5399& 53.1764984&-27.7897053&24.981&  7095.1 $\pm$  29.3&    36&  0.420& [OIII]$\lambda\lambda$4959,5007 \\
               &           &           &      &  9322.7 $\pm$  35.1&    13&  0.421&           H$\alpha$ \\
           5435& 53.1508255&-27.7896385&25.566&  8031.3 $\pm$  41.2&     8&  0.224&           H$\alpha$ \\
           5491& 53.1956520&-27.7877655&22.191&  5970.6 $\pm$  25.1&   682&  0.195& [OIII]$\lambda\lambda$4959,5007 \\
               &           &           &      &  7892.5 $\pm$  48.2&   297&  0.203&           H$\alpha$ \\
           5569& 53.1471786&-27.7884827&23.317&  6185.3 $\pm$  25.0&    23&  0.660&  [OII]$\lambda$3727 \\
               &           &           &      &  8254.0 $\pm$  25.9&    60&  0.652& [OIII]$\lambda\lambda$4959,5007 \\
    5606$^\dag$& 53.1421242&-27.7866974&20.719&  8229.4 $\pm$   0.0&   214&  0.254&           H$\alpha$ \\
           5620& 53.1815414&-27.7879868&23.327&  6172.1 $\pm$  35.4&    42&  0.236& [OIII]$\lambda\lambda$4959,5007 \\
               &           &           &      &  8044.7 $\pm$  16.4&    52&  0.226&           H$\alpha$ \\
           5815& 53.1983261&-27.7878132&25.955&  9426.5 $\pm$  29.6&    12&  0.887& [OIII]$\lambda\lambda$4959,5007 \\
           5933& 53.1782417&-27.7872906&25.379&  6742.2 $\pm$  31.6&    23&  0.350& [OIII]$\lambda\lambda$4959,5007 \\
               &           &           &      &  8836.6 $\pm$  22.5&    10&  0.346&           H$\alpha$ \\
           5959& 53.1640968&-27.7872963&24.325&  7831.5 $\pm$  27.3&    33&  0.193&           H$\alpha$ \\
           6022& 53.1471252&-27.7871094&25.249&  8891.0 $\pm$  26.5&     3&  0.780& [OIII]$\lambda\lambda$4959,5007 \\
           6082& 53.1447182&-27.7854404&22.470&  6131.5 $\pm$  13.4&   953&  0.228& [OIII]$\lambda\lambda$4959,5007 \\
               &           &           &      &  8177.0 $\pm$  16.7&   392&  0.246&           H$\alpha$ \\
           6139& 53.1581268&-27.7863865&25.559&  7145.9 $\pm$  63.1&    22&  4.877&          Ly$\alpha$ \\
           6162& 53.1634712&-27.7866364&25.340&  7388.6 $\pm$  27.8&    16&  0.479& [OIII]$\lambda\lambda$4959,5007 \\
           6196& 53.1604767&-27.7862968&24.792&  7387.6 $\pm$  29.3&     4&  0.982&  [OII]$\lambda$3727 \\
           6710& 53.1925850&-27.7838020&25.017&  9301.4 $\pm$  32.7&     4&  0.862& [OIII]$\lambda\lambda$4959,5007 \\
           6732& 53.1784897&-27.7840405&24.625&  6494.7 $\pm$  48.6&   137&  3.193&  [CIV]$\lambda$1549 \\
               &           &           &      &  7991.4 $\pm$  22.8&    25&  3.186& [CIII]$\lambda$1909 \\
           6846& 53.1845741&-27.7833309&25.193&  7990.1 $\pm$  24.4&    16&  1.144&  [OII]$\lambda$3727 \\
           6853& 53.1518402&-27.7828617&23.428&  6964.1 $\pm$  26.3&    24&  0.869&  [OII]$\lambda$3727 \\
               &           &           &      &  9280.7 $\pm$  36.9&    55&  0.858& [OIII]$\lambda\lambda$4959,5007 \\
           6893& 53.1927605&-27.7835293&25.334&  7180.5 $\pm$  41.1&    28&  0.438& [OIII]$\lambda\lambda$4959,5007 \\
           6953& 53.1527824&-27.7826958&24.721&  6609.8 $\pm$  37.5&    10&  0.774&  [OII]$\lambda$3727 \\
               &           &           &      &  8812.3 $\pm$  41.0&    53&  0.764& [OIII]$\lambda\lambda$4959,5007 \\
           6959& 53.1765175&-27.7825508&25.051&  6636.4 $\pm$  21.3&    38&  0.329& [OIII]$\lambda\lambda$4959,5007 \\
               &           &           &      &  8784.3 $\pm$  25.3&    21&  0.338&           H$\alpha$ \\
           7131& 53.1352577&-27.7816753&24.314&  9121.6 $\pm$  33.3&     8&  1.447&  [OII]$\lambda$3727 \\
           7398& 53.1763268&-27.7808590&24.892&  7576.1 $\pm$  27.2&    50&  0.517& [OIII]$\lambda\lambda$4959,5007 \\
           7687& 53.1505852&-27.7712173&23.030&  6284.6 $\pm$   9.1&    23&  0.258& [OIII]$\lambda\lambda$4959,5007 \\
               &           &           &      &  8155.5 $\pm$  35.3&    64&  0.243&           H$\alpha$ \\
           7847& 53.1739960&-27.7720566&21.227&  8821.8 $\pm$  71.6&   231&  0.344&           H$\alpha$ \\
           7889& 53.1844864&-27.7722511&25.113&  9221.3 $\pm$  37.3&    21&  0.846& [OIII]$\lambda\lambda$4959,5007 \\
           8026& 53.1512146&-27.7728405&23.967&  7912.0 $\pm$  24.5&    71&  0.206&           H$\alpha$ \\
           8040\tablenotemark{d}& 53.1619606&-27.7739010&22.532&  6359.6 $\pm$  38.2&    43&  0.273& [OIII]$\lambda\lambda$4959,5007 \\
               &           &           &      &  8388.3 $\pm$  41.8&    29&  0.278&           H$\alpha$ \\
           8040\tablenotemark{d}& 53.1619606&-27.7739010&      &  6308.2 $\pm$  28.0&    59&  0.263& [OIII]$\lambda\lambda$4959,5007 \\
               &           &           &      &  8367.8 $\pm$  56.9&    39&  0.275&           H$\alpha$ \\
    8157$^\dag$& 53.1649513&-27.7736721&28.344&  6407.1 $\pm$   5.5&     6&  0.000&                     \\
           8680& 53.1477776&-27.7769489&23.995&  7780.0 $\pm$  21.4&     4&  1.087&  [OII]$\lambda$3727 \\
           8744& 53.1466179&-27.7774906&25.416&  7819.3 $\pm$  27.1&     5&  0.191&           H$\alpha$ \\
        9040& 53.1711852&-27.7784585&26.183&  7350.9 $\pm$  50.1&    13&   5.045&          Ly$\alpha$            \\
%          9088-old& 53.1463614&-27.7700826&25.277&  7891.7 $\pm$  27.1&    3.2 &   0.580 & [OIII]$\lambda$4995            \\
           9088& 53.1463623&-27.7700825&25.281&  7891.7 $\pm$  27.1&     7&  0.580& [OIII]$\lambda\lambda$4959,5007 \\
           9244& 53.1615181&-27.7676182&23.685&  8380.4 $\pm$  31.1&    41&  0.678& [OIII]$\lambda\lambda$4959,5007 \\
           9340& 53.1694679&-27.7655869&27.517&  6940.5 $\pm$  23.7&    15&  4.708&          Ly$\alpha$ \\
           9397\tablenotemark{e}& 53.1628532&-27.7671661&21.020&  6225.0 $\pm$  34.4&   747&  1.225&  [MgII]$\lambda$2798 \\
               &           &           &      &  7080.8 $\pm$  29.9&   140&       &  FeII complex        \\
               &           &           &      &  9661.6 $\pm$  35.9&    69&       &  H$\gamma$-FeII-[OIII] complex  \\
           9437& 53.1492195&-27.7637062&23.483&  8684.1 $\pm$  54.8&    32&  0.323&           H$\alpha$ \\
           9487& 53.1673927&-27.7668190&27.204&  6194.0 $\pm$  30.7&    17&  4.094&          Ly$\alpha$ \\
           9712& 53.1579247&-27.7552757&25.514&  7142.5 $\pm$  48.5&    17&  0.916&  [OII]$\lambda$3727 \\
               &           &           &      &  9533.8 $\pm$  68.4&    15&  0.909& [OIII]$\lambda\lambda$4959,5007 \\
           9765& 53.1516647&-27.7613010&23.581&  7448.4 $\pm$  27.8&    26&  0.998&  [OII]$\lambda$3727 \\
           9962& 53.1562233&-27.7573833&24.177&  9196.2 $\pm$  34.2&    20&  0.841& [OIII]$\lambda\lambda$4959,5007 \\
          10025& 53.1704445&-27.7613754&21.517&  7641.0 $\pm$  37.8&   131&  0.164&           H$\alpha$ \\
          20037& 53.1593552&-27.7750263&24.665&  8273.6 $\pm$  31.8&    17&  1.220&  [OII]$\lambda$3727 \\
   20039$^\dag$& 53.1604271&-27.7752285&26.154&  9168.0 $\pm$   4.1&    54& -1.000&                     \\
          20061& 53.1624680&-27.7803574&26.029&  8586.0 $\pm$  32.9&    44&  0.719& [OIII]$\lambda\lambda$4959,5007 \\
          20083& 53.1865730&-27.7902279&23.658&  7333.6 $\pm$  19.2&    74&  0.468& [OIII]$\lambda\lambda$4959,5007 \\
\enddata
\vskip 0.5cm
\tablenotetext{a}{Wavelength errors are the formal uncertainty from a gaussian
fit to the line.  Based on comparison of GRAPES redshifts
with VLT redshifts (see text), we estimate that a systematic error 
floor of $\sim 12$\AA\ should be applied when the quoted error
is smaller.}
\tablenotetext{b}{Fluxes quoted here are directly measured from
the emission features falling within the grism extraction window.
Because this window is often relatively narrow, we estimate that
the total emission line flux of an object could exceed the tabulated
fluxes by aperture correction factors in the range of $\sim 1.5$ to $\sim 3$.
}
%\tablenotetext{$\dag$}{These objects are manually added into the 
%emission line list.}
\tablenotetext{\dag}{These objects are manually added into the 
emission line list.}
\tablenotetext{c}{Line flux and redshift for object 5225 are taken from 
Rhoads et al 2005.}
\tablenotetext{d}{Object 8040 in the UDF catalog corresponds to two
distinct catalog entries in an earlier GRAPES catalog, which likely
correspond to a pair of HII regions in the same galaxy.
The two regions were analyzed separately by ``emlinecull,'' resulting
in two table entries for this object.  The total line flux for the
object is approximately the sum of the two entries, while both
have the same redshift within the uncertainties.}
\tablenotetext{e}{Line identifications for object 9397 are
based on comparison with the composite quasar spectra of Francis
et al (1991) and vanden Berk et al (2001).  The redshift is based on 
the MgII line.}
%  \hskip-4.25in $^\dag$ These objects are manually added into the emission line list.
\end{deluxetable}
% \end{document}

%\documentclass{aastex}
%\begin{document}
\begin{deluxetable}{cccc}
\tablewidth{0pt}
\tablecaption{Redshift comparison between GRAPES and ground-based spectra.}
\tablehead{ 
\colhead{UDF}&\colhead{z (GRAPES)}&\colhead{z (VLT)}&
\colhead{reference\tablenotemark{a}}
}

\startdata
  -100 & 0.601  & 0.218,0.6354\tablenotemark{b}  & 1,2 \\
  -101 & 0.138  & 0.132  & 2  \\
  -104 & 1.218  & 1.220  & 4\\
   900 & 0.419  & 0.414  & 2,3 \\
  1000 & 0.232  & 0.213  & 2 \\
  1446 & 2.470  & 2.470  & 5\\
  1752 & 1.238  & 1.244  & 1 \\
  2201 & 1.377  & 1.382  & 4\\
  3484 & 0.405  & 0.947  & 2 \\
  4142 & 0.729  & 0.737  & 1 \\
  4396 & 1.218  & 1.223  & 1 \\
  4445 & 0.922  & 0.458  & 3 \\
  4816 & 1.230  & 1.220  & 1 \\
  4929 & 0.438  & 0.438  & 2,3 \\
  4981 & 1.430  & 1.438  & 4\\
  5606 & 0.254  & 0.229  & 2 \\
  5620 & 0.226  & 0.212  & 2 \\
  6732 & 3.193  & 3.193  & 3 \\
  6953 & 0.764  & 0.765  & 4\\
  7847 & 0.344  & 0.333  & 1,2 \\
  8680 & 1.088  & 1.086  & 4\\
  9397 & 3.017  & 1.216  & 3 \\
 20037 & 1.220  & 1.215  & 4\\
\enddata
\tablenotetext{a}{
The references: (1) Vanzella et al.~2005; 
 (2) Le Fevre et al.~2004; (3) Szokoly et al.~2004;
 (4) Vanzella et al.~2006; (5) Grazian et al.~2006}
\tablenotetext{b}{Object -100 has discrepant ground-based
redshifts.  The database of Le Fevre et al yields $z=0.6354$, in rough
agreement with the GRAPES value, while the database of Vanzella
et al (2005) gives 0.218.}
\end{deluxetable}

\end{document}